\documentclass[12pt]{article}
\usepackage{amsmath}
\usepackage{graphicx,psfrag,epsf}
\usepackage{enumerate}
\usepackage{natbib}
\usepackage{url} % not crucial - just used below for the URL 
\usepackage{pdflscape}

\usepackage{algorithm}

\usepackage{amsfonts}
\usepackage{amsmath}
\usepackage{amssymb}
\usepackage{graphicx}
\usepackage{subfig}
\usepackage{booktabs}
\usepackage[width=1.25\textwidth]{caption}
\usepackage{longtable}
\usepackage{array}
\usepackage{multirow}
\usepackage{wrapfig}
\usepackage{float}
\usepackage{colortbl}
\usepackage{multirow}
\usepackage{pdflscape}
\usepackage{threeparttable}
\usepackage{threeparttablex}
\usepackage[normalem]{ulem}
\usepackage{makecell}
\usepackage{xcolor}
\usepackage{natbib}
\usepackage[nameinlink,noabbrev]{cleveref}
\usepackage{appendix}
\usepackage{algorithm}
\usepackage{algpseudocode}
\usepackage{tikz}

\def\spacingset#1{\renewcommand{\baselinestretch}%
	{#1}\small\normalsize} \spacingset{1}\def\spacingset#1{\renewcommand{\baselinestretch}%
	{#1}\small\normalsize} \spacingset{1}

\usetikzlibrary{shapes.geometric, arrows.meta, positioning, decorations.pathreplacing}

\begin{document}

\title{Coherent Hierarchical Forecasting for Proportion and Discrete Time Series}
\author{Hannah Comiskey \thanks{The author has been
supported by Australian Research Council (ARC) Discovery Project Grant
DP200101414.}}
\maketitle

\noindent\textbf{Acknowledgements.}
The author gratefully acknowledges the contributions of Professor David T. Frazier and Associate Professor Ole Maneesoonthorn.

\medskip

\noindent\textbf{Notice.}
This is a working paper and is not to be circulated. It is for discussion and comment purposes only. This paper has not been peer-reviewed.

\begin{abstract}
    Hierarchical and grouped time series arise when a multivariate time series is forced to satisfy a set of aggregation constraints, motivating forecast reconciliation methods that ensure coherent forecasts of such hierarchical structures. Many real-world applications involve discrete or bounded supports, introducing additional challenges that are not addressed Gaussian-based reconciliation methods. We develop a post-hoc hierarchical forecasting approach to construct coherent forecast hierarchies for discrete and bounded time series. The method constructs coherent forecasts by convolution and exponential tilting, preserving the distributional properties and the underlying support throughout the hierarchy. We evaluate the proposed approach against state-of-the-art reconciliation methods for both discrete and continuous settings, demonstrating strong performance across a range of experiments. Through simulation studies and empirical applications in epidemiological and demographic data, we show that the method provides reliable and coherent distributional forecasts in challenging scenarios. %This approach further advances hierarchical probabilistic forecasting beyond classical Gaussian frameworks, offering a flexible and practically applicable solution for non-Gaussian hierarchical time series.
\end{abstract}
 
\spacingset{1.8} % DON'T change the spacing!

\section{Introduction} \label{introduction}

Multivariate time series often exhibit an underlying grouping structure that allows the individual series to be organised hierarchically as nodes. Parent nodes are formed by aggregating their child nodes, while the most disaggregated series are referred to as bottom-level nodes \citep{Athanasopoulos2020}. It is well established that exploiting the aggregation constraints underlying a hierarchy can improve forecast accuracy compared to forecasting each series independently \citep{Athanasopoulos2024}. This improvement is the result of ensuring forecast coherence. Coherence is a property of data when the aggregation of each series in the hierarchy respects a set of linear constraints \citep{Panagiotelis2021}. If forecasts are generated without imposing the underlying grouping structure, such as when series are generated independently, then the resulting forecasts will not satisfy the hierarchical structure and will not be `coherent'. Incoherent forecasts do not properly use all available information, which can increase the variability of the forecast \citep{Panagiotelis2021, Athanasopoulos2023}. \\
\\
Ensuring forecast coherence in the case of continuous multivariate series for both point- and probabilistic forecasts is now well-understood, see \cite{Hyndman2011, Hyndman2016, Wickramasuriya2019, Jeon2019, Eckert2021, Panagiotelis2023, Wickramasuriya2024} and more recently \cite{Athanasopoulos2024} for a complete review. Point forecasts only consider the expected value, rather than the complete predictive distribution. There are four main types of point-forecast reconciliation methods: The bottom-up (BU) approach which forecasts the bottom series and aggregates upwards, the top-down (TD) approach which forecasts the top series and disaggregates downwards, the middle-out (MO) approach which combines aspects from both the bottom-up and top-down approaches, and optimal reconciliation methods which adjust all base forecasts in the series to minimise a loss function \citep{Hyndman2021}. The BU, TD and MO forecast a single level of the hierarchy and then aggregate or disaggregate accordingly to estimate the remaining structure. This does not account for the multivariate nature of the data. Optimal reconciliation methods propose forecasting each level of the hierarchy first, we will call these the base forecasts. Using a regression framework, the constraints of the system are considered via a regression model. The set of estimates from the regression model are considered the coherent forecasts \citep{Hyndman2011}. Building on this work, the Minimum Trace (MinT) approach considers the problem as an optimisation problem, rather than a regression problem. First \citet{Wickramasuriya2019} assume that the base forecasts are unbiased, and proposes an elegant solution to the issue by minimising the trace of the reconciled forecast error covariance matrix. In recent developments, \citet{Panagiotelis2021} propose a geometric interpretation of the hierarchical forecasting problem. This approach proved why and how reconciliation via projections improves forecast accuracy for a given class of loss functions. \\
\\
Moving beyond point-forecasting, decision-making often requires uncertainty information. Probabilistic forecasting considers the complete predictive distribution, requiring the distributions themselves to be coherent rather than just their point-estimates. Coherency, as it relates to predictive distributions, occurs when the reconciled forecasts satisfy the constraints of the underlying hierarchical structure system \citep{Panagiotelis2023}. \citet{Panagiotelis2023} propose a probabilistic method estimating reconciliation weights which optimise the energy or variogram score over a set of base forecasts via stochastic gradient descent. However, existing probabilistic reconciliation methods are generally developed for continuous variables on an unrestricted support, where linear transformations of the predictive distributions can be readily defined and, in some cases, preserve the assumed distributional family \citep{Wickramasuriya2019, Panagiotelis2023}. Yet for constrained data, such as proportions or counts, assuming an unconstrained probability space can generate incoherent draws outside the feasible range or distort the underlying distributional shape. In such cases, coherence may require forecasts to satisfy minimum or maximum constraints, while higher-level series may also be subject to aggregation constraints. Therefore, the case of bounded or discrete hierarchies is less well-understood and studied. Recent solutions to this problem include the approaches of \citet{Corani2023}, \citet{Zambon2024}, and \citet{ZhangPanagiotelisKang2024} to ensure reconciliation of the resulting forecasts. In \citet{Corani2023}, the authors use a generalised version of Bayes Rule to yield a reconciled probability mass function. \citet{Zambon2024} proposes an algorithm called `Bottom-Up Importance Sampling' (BUIS) to efficiently sample from any class of reconciled probability distribution. Finally, \citet{ZhangPanagiotelisKang2024} proposes the `Discrete Forecasting Reconciliation' (DFR) algorithm which optimises the penalised Brier score. Collectively, these advances establish an important foundation for reconciling full predictive distributions in constrained spaces, and directly motivate the methods developed in this paper. \\
\\
Despite these recent advances, we identify three main issues with probabilistic forecast reconciliation of constrained time series. Firstly, the models used for bottom-series forecasting are likely biased in constrained series. For example, it is well-known that no unbiased estimator of a non-negative population quantity exists, directly invalidating the no-bias assumption that sits at the core of nearly all reconciliation approaches. Secondly, suppose that the bottom-series models are unbiased. Once the parameter uncertainty of these models are accounted for, the aggregated forecast distributions may be intractable, making reconciliation according to the system constraints difficult. Finally, existing reconciliation methods aim to modify the base forecasts across the hierarchy so that all series jointly satisfy the aggregation constraints. While this produces a coherent set of forecasts, it may unnecessarily alter predictive distributions that are already well-calibrated to the characteristics of their respective series. This is particularly relevant when the series differ substantially in their levels of uncertainty or distributional shape. Rather than modifying all base forecasts to achieve coherence, an alternative is to construct coherent forecasts by retaining the information contained in the base predictive distributions and imposing the aggregation constraints only where necessary. \\
\\
To address these challenges, we develop a method for constructing coherent hierarchical forecasts for bounded and discrete multivariate time series. Rather than fully reconciling all base forecasts, our approach constructs a coherent set of predictive distributions while preserving the information contained in the independently generated base forecasts. Specifically, we use convolution to aggregate the base predictive distributions, thereby propagating their distributional uncertainty to higher levels of the hierarchy, and exponential tilting to adjust the resulting distributions so that their predictive moments satisfy the required aggregation constraints. This provides a flexible framework for constructing coherent forecasts while respecting the natural support and distributional characteristics of bounded and discrete variables. We demonstrate its performance using simulated Beta and Poisson hierarchical data structures and real-world empirical applications, benchmarking against established reconciliation methods. \\
\\
The remainder of the paper is organised as follows. Section 2 outlines the proposed methodology, including implementation for Poisson and Beta hierarchies and our evaluation strategy. Section 3 presents simulation study and empirical application used to assess predictive performance of discrete count time series data. Section 4 presents simulation study and empirical application used to assess predictive performance of constrained proportion time series data. Section 5 discusses the advantages and limitations of the approach and highlights opportunities for further extensions.

\section{Coherent Forecasts with Convolution and Tilting}

\subsection{Notation}
To begin, let $m$ be the number of series at the bottom level, and $n$ be the total number of series in the hierarchical set. We denote the bottom series by $\boldsymbol{b}_t$, such that 
$$\boldsymbol{b}_t = [b_{1,t}, b_{2,t}, \dots, b_{m,t}]^\top,$$
where each bottom-level series $b_{i,t}$ takes values in the sample space $\mathcal{Y}_i$. Consequently, the vector of bottom-level series takes values in the product space $\mathbb{R}^m$, with the $\sigma$-algebra $\mathcal{F}_{m}$ and probability measure $\mu_{m}$, forming the probability space $(\mathbb{R}^m, \mathcal{F}_{m}, \mu_{m})$. Similarly, let $\boldsymbol{u}_{t}$ denote the $n_{u}$-dimensional vector of upper-level series observations. Here, $\boldsymbol{u}_{t} = [u_{1,t}, u_{2,t}, \dots, u_{n_{u},t}]$, with $n_{u} \leq m$ and $n_{u} + m = n$. The vector $\boldsymbol{u}_{t}$ take values in the probability space ($\mathbb{R}^{n_{u}}, \mathcal{F}_{u}, \mu_{u}$). Let $\boldsymbol{y}_{t}$ denote the vector of series at time $t$, and $y_{i,t}$ the value at series $i$ and time $t$, $ i \leq n$ and $t \leq T$. Therefore, the hierarchical relationships within $\boldsymbol{y}_{t}$ can be expressed compactly as
\begin{equation}
\boldsymbol{y}_t = \begin{bmatrix}
\boldsymbol{u}_t \\
\boldsymbol{b}_t
\end{bmatrix}  = \boldsymbol{S} \boldsymbol{b}_t,
\end{equation}
where $\boldsymbol{y}_{t} \in \mathbb{R}^{n}$ denotes the vector of all $n$ series at time $t$, and $\boldsymbol{b}_{t} \in \mathbb{R}^m$ denotes the vector of the $m$ bottom-level series. The hierarchical structure is represented by the ${n \times m}$ summing matrix
$ \boldsymbol{S} = \begin{bmatrix} \boldsymbol{S}_{u} \\ \boldsymbol{I}_{m} \end{bmatrix}$.  Here, $\boldsymbol{S}_{u}$ is the ${n_{u} \times m}$ summing matrix that maps the $m$ bottom-level series to the $n_{u}$ upper-level series, and $\boldsymbol{I}_{m}$ is the $m \times m$ identity matrix. The $\boldsymbol{S_{u}}$ matrix can include any real values, specifying the linear constraints of the system \citep{Athanasopoulos2020}. The matrix $\boldsymbol{S}$ can also be considered as a function $s: \mathbb{R}^m \mapsto \mathbb{R}^n$ which links the bottom base bottom series forecasts $\boldsymbol{b_{t}}$ to the coherent vector $s(\boldsymbol{b_{t}}) = \boldsymbol{S} \boldsymbol{b}_t$. As such, the coherent vector $\boldsymbol{y}_t$ exists in the vector subspace $\mathsf{s}$, which is well-defined in $\mathbb{R}^n$ and is spanned by the columns of $\boldsymbol{S}$. A forecast distribution is said to be coherent if the random variables associated with the reconciled forecasts satisfy the same aggregation constraints as the underlying hierarchical structure. According to \citet{Panagiotelis2023}, a probability triple  $(\mathsf{s}, \mathcal{F}_{\mathsf{s}}, \tilde{\mu}_{\mathsf{s}})$ is said to be coherent with the bottom probability triple $(\mathbb{R}^m, \mathcal{F}_{m}, \mu)$ if, 
\begin{align}
  \tilde{\mu}_{\mathsf{s}}(\mathsf{s}(\mathcal{B})) = \mu(\mathcal{B}) \quad \forall \mathcal{B} \in \mathcal{F}_{m}.
\end{align} 
Consequently, the marginal distributions of the upper-level series are induced by the joint distribution of the bottom-level series through the aggregation structure.

\subsection{Forecast convolution}
Let the forecast distribution of bottom-level series $c$ at time $t$ be given by,
\begin{equation*}
    b_{c,t} \sim f_{c,t}(\cdot), \quad 1 \leq c \leq m.
\end{equation*} 
We assume that the series in the hierarchical structure are mutually independent such that,
\begin{align}
    f_{t}(b_{1,t}, \dots, b_{m,t}) &= \prod_{c=1}^{m}f_{c,t}(b_{c,t}).
\end{align}
Under this assumption, aggregation via convolution is directly analogous to a bottom-up aggregation strategy. Let $\mathcal{D}_d$ denote the set of bottom-level descendants of node $d$, where $m_{d}$ = $|\mathcal{D}_d|$ and $u_{d, t} = \sum_{c \in \mathcal{D}_d} b_{c,t}$. Then, the forecast density at time $t$ may be obtained via convolution where, 
\begin{align}
    \label{eqn_contconv}
    f_{d, t}(u_{d, t}) = \int \dots \int \prod_{c=1}^{m_{d}-1}f_{c,t}(b_{c,t})f_{m_{d},t}(u_{d, t} - \sum_{c=1}^{m_{d}-1}b_{c,t})db_{c,t}
\end{align}
Equation \eqref{eqn_contconv} may be approximated using Monte Carlo integration making it computationally simple to evaluate. In practice, the resulting convolution density, $f_{d,t}(u_{d,t})$, can be tilted towards the predictive mean of the base forecast for the aggregate quantity, $\hat{u}_{d,t}$. Since lower-level descendants are often noisier than their parent nodes, the mean implied by the convolution density may differ from the predictive mean obtained from a model tailored directly to the aggregate series. Tilting therefore aligns the distribution obtained through bottom-up aggregation with the information contained in the aggregate-level base forecast.

\subsection{Forecast tilting}
Tilting is a well-established process that allows for the incorporation of additional information into the convoluted density to change the shape of the density, while preserving its support \citep{GiacominiRagusa2014, KruegerClarkRavazzolo2017, West2024}. It incorporates a set of theoretical restrictions, expressed as $k$ moment conditions, to modify the probability distribution via re-weighting the density or mass function. 
Let \begin{align*}
    u_{d,t} \sim f_{d, t}(\cdot),\\
    E(u_{d,t}) = \hat{u}_{d,t}.
\end{align*}
We tilt the convoluted density, $f_{d,t}(u_{d,t})$, towards an updated density $f^{*}_{d, t}(d, t)$, via matching the first moment $\hat{u}_{d,t}$.

Therefore,
\begin{align}
    f_{d, t}^{*}(u_{d, t}) &= \frac{ \text{exp}(\gamma_{d,t} u_{d,t}) f_{d,t}(u_{d,t})}
    {\int \text{exp}(\gamma_{d,t} u_{d,t}) f_{d,t}(u_{d,t}) du}, 
\end{align}
where $\gamma_{d,t}$ is the tilting parameter that minimises the residual between the target mean and the mean of the tilted density for series $d$ at time $t$. A complete description of the estimation process for $\gamma_{d,t}$ is given in the Appendix, section \ref{appendix_tilting_details}. This tilted density is approximated by numerical integration \citep{BurdenFaires2015}. \\
\\
When a closed-form expression for the tilted predictive density is available, tilting parameters can be estimated sequentially for each parent node in a hierarchical structure. In such settings, the predictive density at a parent node can be constructed by aggregating the predictive densities of its child nodes, and the corresponding tilting parameter can be obtained via the moment-matching procedure described above. This sequential approach proceeds up the hierarchy, using the aggregated and tilted densities at lower levels as the starting point for higher-level nodes. Estimating tilting parameters sequentially, rather than independently across all series, increases the effective degrees of freedom and provides greater flexibility in the resulting predictive densities. When a closed-form expression for the tilted predictive density is unavailable, we recommend tilting each series independently. Specifically, convolutions are used to obtain the predictive density of each parent series. This density is then tilted towards the predictive mean of the corresponding base forecast, thereby incorporating information from the model fitted directly to the parent node.

\subsection{Tilting through the hierarchy}
The approach used to tilt the predictive densities depends on whether or not the tilted density is available in closed form. Algorithm ~\ref{algorithm_tilt} describes both approaches. Appendix \ref{appendix_tilting_details} describes the process for estimating the tilting parameter $\gamma_{d, t}$. \\
\\
When a titled density is available in closed form, the tilting procedure can be applied sequentially through the hierarchy. Starting at the bottom-level nodes, each aggregated node is constructed from the predictive densities of it's children. The tilting procedure is applied such that the first moment of the aggregated predictive density matches that of the target parent node. This tilted predictive density is then used in the construction of subsequent predictive densities of parent nodes. This procedure is repeated until the predictive density of the top-level node is estimated and tilted.\\
\\
In contrast, when no closed-form expression for the convoluted parent density is available, each series must be tilted independently. In this instance, the predictive density of each parent node is generated from the convolution of child predictive densities. The resulting density is then tilted toward the target predictive node of the corresponding parent node. As the tilted density cannot be directly propagated into the subsequent convolutions, the tilting procedure is repeated for each series independently.

\begin{algorithm} [H]
\caption{Forecast tilting through a hierarchy}
\label{algorithm_tilt}
\begin{algorithmic}[1]
\Require Base predictive densities $f_{d, t}(\cdot)$ and target means $\hat{u}_{d, t}$ for all series $d$
\For{d in ($m+1, \dots, n$)}
    \State Construct the predictive density $f_{d, t}(\cdot)$ via convolution of the predictive densities of its child nodes.
    \If{the tilted predictive density is available in closed form}
        \State Estimate the tilting parameter $\gamma_{d, t}$ such that $E_{f^{*}_{d, t}}(u_{d, t}) = \hat{u}_{d, t}$.
        \State Obtain the tilted predictive density $ f^{*}_{d, t}(u_{d, t}) = \frac{\exp(\gamma_{d, t}u_{d, t})f_{d, t}(u_{d, t})}{\int \exp(\gamma_{d, t}u_{d, t})f_{d, t}(u_{d, t})\,du}$.
        \If{$d < n$}
        \State Use $f^{*}_{d, t}(\cdot)$ as an input for aggregation via convolution for $d+1$.
        \EndIf
    \Else
        \State Estimate the tilting parameter $\gamma_{d, t}$ such that $ E_{f^{*}_{d, t}}(u_{d, t}) = \hat{u}_{d, t}$.
        \State Obtain $f^{*}_{d, t}(\cdot)$ by numerical integration.
    \EndIf
\EndFor
\State \Return $f^{*}_{d, t}(\cdot)$

\end{algorithmic}
\end{algorithm}

\subsection{Evaluation metrics}
In the following simulations and examples, we withheld a portion of the data during model building to generate a rolling window of one-step-ahead forecasts. The forecasts of the withheld observations produced at each iteration were then used to evaluate and compare coherent probabilistic forecasts using strictly proper scoring rules. Scoring rules are described as `strictly proper' when $E_{Q}[R(Q,y)] \leq E_{Q}[R(F,y)]$ where $F$ is any predictive distribution, $Q$ is the true distribution and $y$ is an observed outcome \citep{gneiting2007}. Scoring rules assign negatively orientated values to predictive forecasts based on the observed actual outcome. Reported scores represent the average performance across the validation set. \\
\\
For continuous distributions, we utilise the energy score and continuous ranked probability score (CRPS) \citep{gneiting2007}. In this study, we use the \texttt{scoringRules} package in \texttt{R} to calculate these scoring rules \citep{scoringRules}. The CRPS measures the average squared difference between the predictive cumulative distribution function (CDF) $\hat{F}$ and the empirical CDF of the observation, $F$. By considering the complete probability distribution rather than just point estimates, the CRPS provides a complete picture of method performance. It is given by,
\begin{align*}
\text{CRPS}(F, y_{i, t}) = \int_{-\infty}^{\infty} \big( \hat{F}(z) - \mathbb{I}\{y_{i, t} \le z\} \big)^2 \, dz.
\end{align*}
The discrete analogue of the CRPS is the discrete ranked probability score (DRPS) \citep{SnyderOrdBeaumont2012}. DRPS uses the L-2 norm to measure the distance between the two distributions and is given by, 
\begin{align*}
\text{DRPS}(\hat{F}, y_{i, t}) = \sum_{z=0}^{\infty} \big( \hat{F}(z) - \mathbb{I}\{y_{i, t} \le z\} \big)^2 .
\end{align*}
The energy score generalises both the CRPS and DRPS. It measures how close a forecasts predicted distribution is, on average, to the true distribution of the data.
\begin{align*}
\text{ES}(F, \boldsymbol{y}_{t}) = \mathbb{E}_F \| X - \boldsymbol{y}_{t} \| - \frac{1}{2}\mathbb{E}_F \| X - X' \|,
\end{align*}
where $X$ and $X'$ are independent draws from the predictive distribution $F$.\\
In general, strictly proper scoring rules are defined as expectations with respect to the predictive distribution. For continuous forecasts these expectations are expressed as integrals, whereas for discrete predictive distributions the same expressions reduce to finite sums over the support of the distribution. The Brier score is a strictly proper score for probabilistic classification of discrete distributions \citep{Brier1950}. It measures the accuracy of probabilistic forecasts by measuring the mean squared difference between predicted probabilities ($f_{t}$) and observed outcome ($q_{t}$), such that
\begin{align*}
   \text{BS} = \frac{1}{N} \sum_{t=1}^{N}(f_{t} - q_{t})^2.
\end{align*}
This approach penalises densities that are incorrectly calibrated and can be interpreted like the mean squared error.

Finally, to compare the forecast accuracy of different reconciliation methods, we will utilise a pairwise Diebold-Mariano (DM) test \citep{DieboldMariano1995, harvey1997, Diebold2012}. Let $(\boldsymbol{y}_{t, 1}, \boldsymbol{y}_{t, 2})$ be the pairs of $h$-step ahead forecasts produced by two competing coherent forecasting methods, for t = $1, \ldots, T$. The quality of the forecasts are judged using the specified loss function $d_{t}(\cdot)$. In this study, we consider the energy score for joint predictive distributions. The difference in energy score is given by,
\begin{align*}
    d_{t}(F, \boldsymbol{y}_{t, 2}, \boldsymbol{y}_{t, 2}) = \text{ES}(F, \boldsymbol{y}_{t, 1}) - \text{ES}(F, \boldsymbol{y}_{t, 2}) %g(\boldsymbol{y}_{t, 1}) - g(\boldsymbol{y}_{t, 2}).
\end{align*}
Using this difference function, the test statistic of the DM test is given by, 
\begin{align*}
    R &= [ \hat{V}(\bar{d})]^{-\frac{1}{2}} \bar{d}, \\
   V(\bar{d}) &= T^{-1}\Big[ \kappa_{0} + 2 \sum_{k=1}^{h-1}\kappa_{k} \Big],
\end{align*}
Where, $\kappa_{k}$ is the $k^{th}$ autocovariance of $d_{t}$ and is an estimable quantity.
Under the null hypothesis of the DM test, the asymptotic distribution of this statistic has a standard Normal distribution.

% Entropic tilting chooses a new density by minimising the Kullback-Leibler (KL) divergence from the baseline density while satisfying the constraints, 
% \begin{align}
%     \text{min}_{D*(y^{top})} KLIC(f_{i,t}^{*}(y_{i,t}_{t} \mid \cdot), f_{i,t}(y_{i,t}_{t} \mid \cdot)), \\
%     \text{ subject to } \mathbb{E}_{t}(g(y_{i,t}_{t} )) = 0. \nonumber
% \end{align}
% The solution is also exponential in form (Equation \ref{eqnExptilt}), but the optimization is typically regularized (e.g., via convex optimization with soft constraints or penalized entropy) (\cite{tallman2022}). In Entropic tilting it is possible to have multiple constraints over which you are optimising the Kullback-Leibler (KL) divergence to stay as close to the starting distribution as possible.

% \section{Implementation via examples}
\section{The case of discrete count time series}
Hierarchical structures may naturally arise in count settings. For example, counts at lower-levels may aggregate to form counts at higher levels. In this setting, we illustration the utility of our method when applied to hierarchical count time series data. Suppose each node of a hierarchical structure is modelled with a Poisson distribution. Therefore, it is possible to estimate the tilting parameters for all parent nodes sequentially by exploiting the closed-form expressions of exponentially tilted Poisson densities. As the sum of independent Poisson random variables is also Poisson, we may use a sequential procedure in which lower-level tilted densities are convolved to form the starting density of their parent node. 

\subsection{Generating and tilting convolutions under discrete constraints.} 
\paragraph{Bottom level estimation.} Consider one has available bottom-level forecasts, $b_{c, t} \mid \lambda_{c, t} \sim \mathrm{Poisson}(\lambda_{c, t})$ and let the expected value of these forecasts be given by $\hat{\lambda}_{c, t}$. 
\paragraph{Mid-level convolution and tilting.}
Assume that the $h$-step ahead forecasts for the mid-level node $u_{d, t}$, where $d < n$, and the corresponding predictive mean $\hat{\lambda}_{d, t}$ are available. Let $\mathcal{D}_{d}$ denote the set of bottom nodes associated with the parent node $u_{d, t}$. The expected value of the convoluted density is given by $\sum_{c \in \mathcal{D}_{d}}\lambda_{c, t}$. The estimated PMF of $u_{d, t}$ via convolution is given by,
\begin{align}
    f_{d, t}(u_{d, t}) &=
    \frac{ (\sum_{c \in \mathcal{D}_{d}}\lambda_{c, t})^{u_{d, t}} e^{-(\sum_{c \in \mathcal{D}_{d}}\lambda_{c, t})}}{u_{d, t}!},
    \qquad u_{d, t} \in \mathbb{N}_0.
\end{align}
To align the mean of the convolution with the predictive mean $\hat{\lambda}_{d, t}$, we apply sequential exponential tilting, exploiting the fact that the probability mass function of the tilted distribution is available in closed form. The tilted probability mass function is given by,
\begin{equation}
f^{*}_{d, t}(u_{d, t} = l) =
\frac{ (\exp \gamma_{d, t}\, l)\, f_{d, t}\!\left(u_{d, t} = l\right)}
{\displaystyle \sum_{r=0}^{L} \exp\!\left(\gamma_{d, t}\, r\right)\, f_{d, t}\!\left(u_{d, t} = r\right)}, \quad l \in [0, \dots, L].
\end{equation}
Where, $L$ is a reasonable known upper limit of the mid-series. The expectation of the mid-level series tilted density is given by,
\begin{align}
    & E_{f_{d, t}^{*}}(u_{d, t}) = \hat{\lambda}^{*}_{d, t} =  \frac{ \sum_{l=0}^{L}\exp(\gamma_{d, t} l) \, f_{d, t}(u_{d, t} = l)} {\sum_{r=0}^L \exp(\gamma_{d, t} r) \, f_{d, t}(u_{d, t} = r)}
    \approx \hat{\lambda}_{d, t}, \nonumber
\end{align}

%%\om{Is there a big difference between middle-level and top-level definitions? If you define the hierarchy generically, you would not need two separate descriptions, they both should fit within the general notation.}

\paragraph{Top-level convolution and tilting.} 
We assume that forecasts for the top series $u_{n_{u}, t}$ are available with their corresponding predictive mean $\hat{\lambda}_{n_{u},t}$. Let $\mathcal{E}_{n_{u}}$ denote the set of mid-level nodes associated with the top-level node $u_{n_{u}, t}$. In the case of the Poisson distribution, the tilted mean parameter can be extracted in closed-form. Let $u_{n_{u},t}$ be the aggregate of the tilted mid-level series, such that the PMF for the convolution is given by,
\begin{align}
    f_{n, t}(u_{n_{u}, t}) &=
    \frac{ (\sum_{d \in \mathcal{E}_{n_{u}}}\hat{\lambda}^{*}_{d, t})^{u_{n_{u}, t}} e^{-(\sum_{d \in \mathcal{E}_{n_{u}}}\hat{\lambda}^{*}_{d, t})}}{u_{n_{u}, t}!},
    \qquad u_{n_{u}, t} \in \mathbb{N}_0.
\end{align}
Once more, sequential tilting is applied using using the estimated tilting parameter $\gamma_{n_{u}, t}$. The tilted probability mass function is given by,
\begin{align}
    f^{*}_{n, t}(u_{n_{u}, t} = v)
      &= \frac{\exp(\gamma_{n_{u}, t} v) \, f_{n, t}(u_{n_{u}, t} = v)}
              {\sum_{r=0}^V \exp(\gamma_{n_{u}, t} r) \, f_{n, t}(u_{n_{u},t} = r)}, \quad v \in [0, \dots, V].
\end{align}
Where, $V$ is a reasonable known upper limit of the top-series. The expectation of the top-level series tilted density is given by,
\begin{align}
    & E_{f_{n, t}^{*}}(u_{n_{u}, t}) = \hat{\lambda}^{*}_{n_{u}, t} =  \frac{ \sum_{l=0}^{L}\exp(\gamma_{n_{u}, t} l) \, f_{n, t}(u_{n_{u}, t} = l)} {\sum_{r=0}^L \exp(\gamma_{n_{u}, t} r) \, f_{n, t}(u_{n_{u}, t} = r)} \approx \hat{\lambda}_{n_{u}, t}, \nonumber
\end{align}

\subsection{Simulation 1: Aggregating count forecasts}
\paragraph{Data description.}
The aim of this simulation is to demonstrate applying the proposed method to a simple two-level hierarchy of low-count Poisson distributions. We generate a hierarchical Poisson dataset with $M=4$ bottom-level series and $N=2500$ observations respectively. 
The bottom-level series are constructed using a generalised linear model with an auto-regressive term and quadratic log-link. For each series $m=1,\dots,M$, regression coefficients are given by,
\begin{align*}
\theta_{m} \sim \text{Uniform}(0.4,\,0.5), \quad \nu_{m,t} \sim N(0, 0.1^2)
\end{align*}
The Poisson rate parameter is then defined as
\begin{align*}
\lambda_{m,t} &= \exp\!\big( \theta_{m} + 0.7*\text{log}(\lambda_{m,t-1}) + \nu_{m,t} \big),
\end{align*}
with observed counts generated according to
\begin{align*}
b_{m,t} \mid \lambda_{m,t} &\sim \text{Poisson}(\lambda_{m,t}).
\end{align*}
To introduce additional variability, integer-valued shocks 
$\varepsilon_{m,t} \in \{-1,0,1\}$ are added independently to each $b_{m,t}$. 
The distribution of $\varepsilon_{i,m}$ is series-specific to ensure that noise does not systematically cancel out when aggregating across series:
\begin{align*}
\begin{aligned}
\varepsilon_{1,t} &\sim \{-1,0,1\} \;\; \text{with probabilities } (0.4, 0.4, 0.2), \\
\varepsilon_{2,t} &\sim \{-1,0,1\} \;\; \text{with probabilities } (0.2, 0.7, 0.1), \\
\varepsilon_{3,} &\sim \{-1,0,1\} \;\; \text{with probabilities } (0.33, 0.34, 0.33), \\
\varepsilon_{4,t} &\sim \{-1,0,1\} \;\; \text{with probabilities } (0.2, 0.5, 0.3).
\end{aligned}
\end{align*}
To ensuring non-negativity, the perturbed bottom-level series are then given by
\begin{align*}
b^{*}_{m,t} = \max\!\big(0,\, b_{m,t} + \varepsilon_{m,t}\big).
\end{align*}  
At the aggregate level, the series is obtained by summing across the unperturbed bottom-level counts:
\begin{align*}
u_t  = \sum_{m=1}^M b_{m,t}.
\end{align*}
The final noisy dataset consists of the covariate $x$, the aggregate series $u$, and the four perturbed bottom-level series $(b^{\ast}_{1},\dots,b^{\ast}_{4})$.

\paragraph{Modelling the base forecasts.} 
We utilise an auto-regressive order (1) model structure for both the bottom and top series. For node $y_{i,t}$ where $i \in (1, ... , n )$ at time $t+h$, we assume that,
\begin{align*}
    \lambda_{i, t} &= \text{exp}(\theta_{i} + \phi_{i}*\text{log}(\lambda_{i, t-1})) \\
     y_{i,t} \mid \lambda_{i, t} &\sim \text{Poisson}(\lambda_{i, t})
\end{align*}
where, $\theta_{i}$ is the intercept, $\phi_{i}$ is the auto-regressive term. We estimate the parameters of the linear model using the \texttt{tsglm} function of the \texttt{tscount} package in \textsf{R} \citep{Liboschik2017}. \\
\\
The data was split into a training (2000 observations) and test (500 observations) sets. The training set is used to build the models, while the test set is used to assess the predictive power and reliability of the hierarchical coherent forecast method. Thus, the coherent forecasts are evaluated on on their predictive performance of the test set. We compared the predictive distributions produced by bottom-up importance sampling (BUIS) \citep{bayesrecon} to use as a benchmark comparison of discrete forecast reconciliation. 

\paragraph{Evaluation of simulation study.} 
We evaluate each method using the average DRPS, Energy Score, and Brier Score over the 500 reconciled test observations, together with a pairwise Diebold-Mariano (DM) test (Table \ref{tab:relative_to_BUIS_pois_demo}). We do not consider the performance of the marginal predictive distributions of the bottom-levels in Table \ref{tab:relative_to_BUIS_pois_demo}, as the proposed method does not alter these series. Instead, we extract the top series from the hierarchical tree and evaluate the marginal forecast distribution separately using the DRPS and Brier Score. We then assess joint forecast distribution performance across the complete hierarchy using the Energy Score. In this simulation, the convolution-based methods perform similarly to the BUIS benchmark, suggesting that the proposed methods achieve comparable forecast performance to BUIS. Finally, we examine differences in forecast accuracy between the methods using a pairwise DM test. The average size of the differential for convolutions and BUIS was calculated as $-0.051$, with $51.4\%$ of energy score differences being less than 0. Similarly, for convolutions with tilting and BUIS the average difference was calculated as $-0.027$ with $44.8\%$ of differences being less than 0. While the proposed methods have very similar forecast scores to BUIS, at the 5\% significance level the DM tests indicate that these small differences in joint predictive accuracy are statistically significant.

\begin{table}[htbp]
\centering
\caption{Probabilistic forecast performance comparison relative to the BUIS method for different reconciliation methods across three scoring rules (DRPS, Brier Score, and Energy Score) using a test set of 500 observations. Lower values indicate better performance for the scoring rules. Pairwise Diebold-Mariano (DM) tests compare forecast accuracy between each reconciliation method and the BUIS method, using the difference in Energy Scores of the joint distributions as the loss differential. The null hypothesis is equal predictive accuracy. Asterisks denote statistically significant differences from BUIS according to the DM test (* $p<0.05$, ** $p<0.01$). In the method column, `Convolution' represents the density estimated using convolution only, `BUIS' represents the bottom-up importance sampling method, and `Convolutions with tilting' represents the density estimated using convolution and exponential tilting. Top marginal refers to the marginal distribution of the top series, while joint distribution refers to the joint distribution of the complete hierarchical series.}
\label{tab:relative_to_BUIS_pois_demo}
\small
\begin{tabular}{lrrrr}
\toprule
\multirow{2}{*}{Method} &
\multicolumn{2}{c}{\shortstack{Top \\ marginal}} &
\multicolumn{1}{c}{\shortstack{Joint \\ distribution}} &
\multicolumn{1}{c}{DM test} \\
\cmidrule(lr){2-3}
\cmidrule(lr){4-4}
\cmidrule(lr){5-5}
& DRPS & Brier & Energy & $p$-value \\
\midrule
BUIS
& 1.000 & 1.000 & 1.000 & -- \\
Convolution
& 0.970 & 0.991 & 0.990 & $4.3 \times 10^{-9}$** \\
\shortstack{Convolutions \\ with tilting}
& 0.963 & 0.991 & 0.994 & $0.008112$** \\
\bottomrule
\end{tabular}
\end{table}

\subsection{Application to predicting syphilis cases in the USA}
\begin{figure}[H]
\centering
\includegraphics[width=14cm]{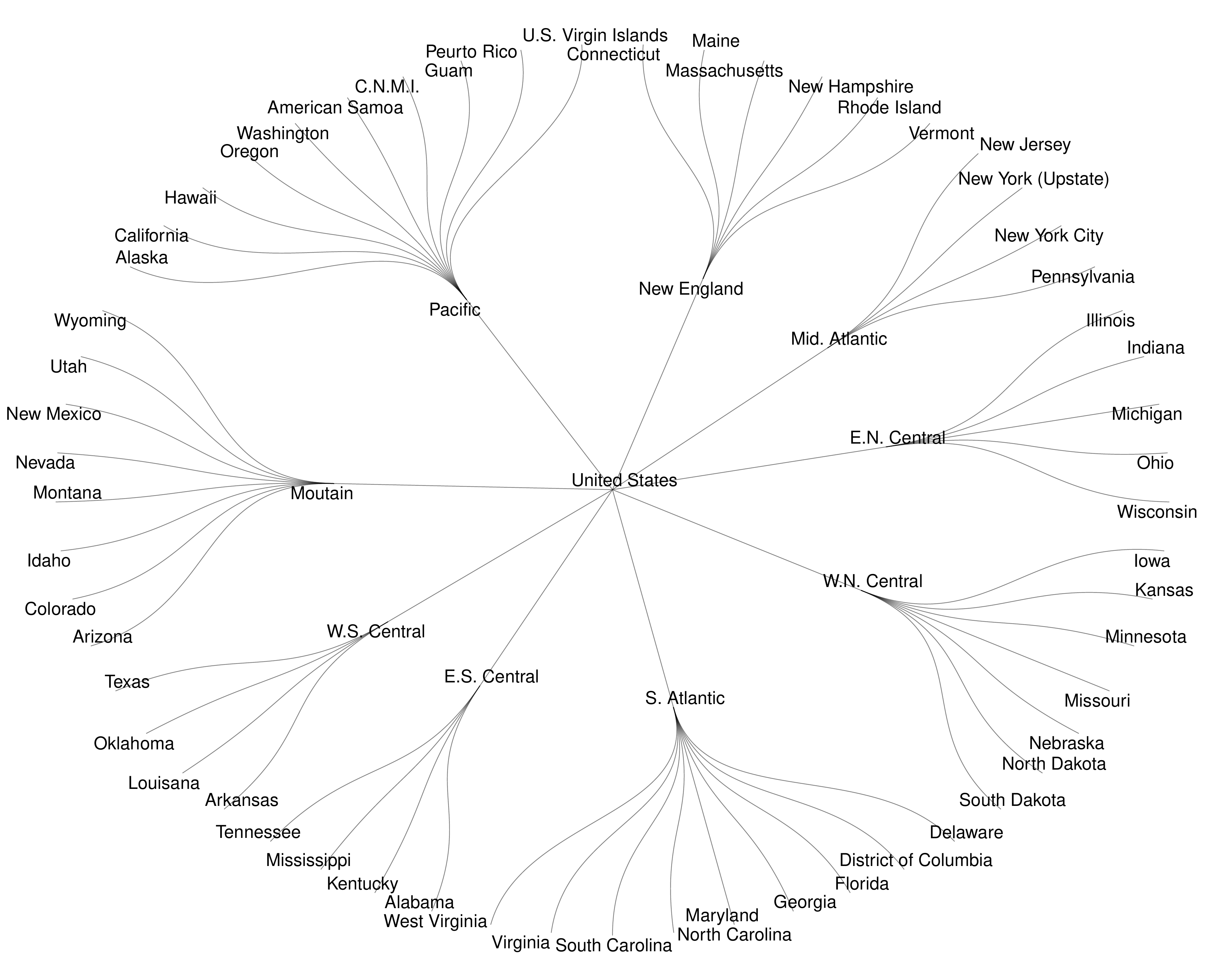}
\caption{The radial dendrogram of the U.S. regional hierarchy. The visualisation displays the hierarchical relationships between national, regional, and state-level nodes. The top node of the series, United States, is at the centre and bottom nodes the capturing the state-level counts are on the outside.}
\label{fig_syph_tree}
\end{figure}

\paragraph{Data description}
The \texttt{syph} dataset from the \texttt{ZIM} package in R contains the weekly number of syphilis cases in the United States from 2007 to 2010 \citep{ZIM}. The data are structured as a hierarchical time series with three levels: the bottom level represents counts by state, the intermediate level aggregates counts by region, and the top level corresponds to total counts across the United States (Figure \ref{fig_syph_tree}). The data contains low-moderate counts at the state-level (range: 0 - 153) with a high proportion of zeroes recorded in the data overall. The highest count recorded is 239 at the national-level.

% Place figure captions after the first paragraph in which they are cited.

\paragraph{Modelling the base forecasts.} 
For a given node in the hierarchical structure $y_{i, t}$ and forecast horizon $t+h$, we assume the observation comes from a Poisson auto-regressive model given by,
\begin{align*}
     \lambda_{i, t} &= \exp(\theta_{i} + \sum_{k=1}^{2} \phi_{i, k}*\text{log}(\lambda_{i, t-1})), \\
     y_{i, t} \mid \lambda_{i, t} & \sim \text{Poisson}(\lambda_{i, t}),
\end{align*}
where, $\theta_{i}$ is the intercept, $\phi_{i, k}$ captures the auto-regressive dependence over the previous 2 weeks. We estimate the parameters of the linear model using the \texttt{tsglm} function of the \texttt{tscount} package in \textsf{R} \citep{tscount}. \\
The data was split into a training and test set. The training set of 183 observations (first 4.5 years of weekly data) is used to build the models, while the test set of 26 observations (final 6 months of weekly data) is used to assess the predictive power and reliability of the hierarchical coherent forecast method. 
\paragraph{Evaluation of weekly syphilis cases in the USA.} 
In this application, the hierarchy comprises three levels; accordingly, we examine both independent and sequential tilting approaches. Table \ref{tab:relative_to_BUIS_pois_syph} shows that the convolution-alone method performs worse than the BUIS benchmark across the performance of the top-marginal and joint predictive distributions, suggesting that this approach alone is insufficient to produce reliable coherent hierarchical forecasts. Furthermore, the DM test considering the difference in energy scores produced by the convolutions and that of the BUIS indicates that a difference may exist between the accuracy of these forecasts. As BUIS is our benchmark, we can conclude that the forecasts produced by convolutions-alone are unreliable. In contrast, both the convolutions with independent and sequential tilting methods achieve comparable forecast performance to BUIS across the top marginal and joint forecast predictive distributions scores, and the DM test failing to prove that a significant difference exists between the energy score of either method and the BUIS benchmark at the 5\% significance level. These results demonstrate that coherent forecasts can be constructed through convolution and tilting without requiring full reconciliation of the base predictive distributions, while retaining competitive probabilistic forecast performance.

\begin{table}[htbp]
\centering
\caption{Probabilistic forecast performance comparison relative to the BUIS method for different reconciliation methods across three scoring rules (DRPS, Brier Score, and Energy Score) using a test set of 26 observations per node. Lower values indicate better performance for the scoring rules. Pairwise Diebold-Mariano (DM) tests compare forecast accuracy between each reconciliation method and the BUIS method, using the difference in Energy Scores of the joint distributions as the loss differential. The null hypothesis is equal predictive accuracy. Asterisks denote statistically significant differences from BUIS according to the DM test (* $p<0.05$, ** $p<0.01$). In the method column, `Convolution' represents the density estimated using convolution only, `BUIS' represents the bottom-up importance sampling method, `Convolutions with ind. tilting' represents the density estimated using convolution and independent exponential tilting and `Convolutions with seq. tilting' represents the density estimated using convolution and sequential exponential tilting. Top marginal refers to the marginal distribution of the top series, while joint distribution refers to the joint distribution of the complete hierarchical series.}
\label{tab:relative_to_BUIS_pois_syph}
\small
\begin{tabular}{lrrrr}
\toprule
\multirow{2}{*}{Method} &
\multicolumn{2}{c}{\shortstack{Top \\ marginal}} &
\multicolumn{1}{c}{\shortstack{Joint \\ distribution}} &
\multicolumn{1}{c}{DM test} \\
\cmidrule(lr){2-3}
\cmidrule(lr){4-4}
\cmidrule(lr){5-5}
& DRPS & Brier & Energy & $p$-value \\
\midrule
BUIS
& 1.000 & 1.000 & 1.000 & -- \\
Convolution
& 5.001 & 0.572 & 3.255 & ${7.152 \times 10^{-10}}^{**}$ \\
\shortstack{Convolutions \\ with ind. tilting}
& 1.040 & 1.03 & 0.992 & 0.7341 \\
\shortstack{Convolutions \\ with seq. tilting}
& 1.040 & 0.987 & 0.994 & 0.8225 \\
\bottomrule
\end{tabular}
\end{table}

\section{The case of time series of proportions}

Proportions and compositional data can be modelled using a range of approaches. Some popular approaches include utilising transformations such as the logit link with a Normal or Multivariate Normal regression \citep{comiskey2025}, Beta regression \citep{ferrari2004beta} and Dirichlet regression \citep{tsagris2018dirichlet}. Here, we focus here on the case where the predictive distributions of the bottom-level proportions are represented by Beta distributions. We present an example to illustrate independent tilting, where no information from previously tilted lower-level series is propagated through to higher levels. This is because when convolving Beta distributions, no closed-form for the joint distribution exists. In this simple example, assume a two-tier hierarchy with bottom-series, $\boldsymbol{b}_t \in (0,1)^m$, parent node $u_{n_{u},t} \in (0,1)$. The complete set of series in the hierarchy is denoted by $\boldsymbol{y}_{t} = (u_{n_u,t}, \boldsymbol{b}_{t})^\top$. The base forecast proportions $\hat{\boldsymbol{b}}_t = (\hat{b}_{1,t}, \dots, \hat{b}_{m,t})^\top$, with corresponding estimated shape parameters $\hat{\boldsymbol{\alpha}}_t = (\hat{\alpha}_{1,t}, \dots, \hat{\alpha}_{m,t})^\top$ and $\hat{\boldsymbol{\beta}}_t = (\hat{\beta}_{1,t}, \dots, \hat{\beta}_{m,t})^\top$ are available. 

\subsection{A remark on using MinT on the logit scale and back-transforming to [0,1].} One possibility for reconciling proportions is to transform them onto the logit scale and assuming Normality on the transformed scale. Removing the bounded nature of the random variable allows the user to utilise MinT methods to reconcile the data, and then back-transform the reconciled estimates onto the [0,1] scale. A key assumption of the MinT method is that the conditionally stationary base forecast errors are unbiased \citep{Wickramasuriya2019}. However, in the Appendix Section \ref{InvLogit_MinT_Bias_Proof}, we show that this assumption no longer holds after back-transformation. The non-linearity of the inverse-logit function induces bias, even when the reconciled forecasts are unbiased on the logit scale. Specifically, the induced bias exhibits an approximately quadratic relationship with the reconciled mean on the probability scale. The magnitude of this bias increases as the reconciled proportion approaches $0.5$, where the curvature of the inverse-logit function is greatest. Conversely, for proportions near the boundaries of the unit interval, the curvature is smaller and the back-transformed MinT reconciled forecasts are approximately unbiased. Therefore, although logit-scale reconciliation provides a convenient way to apply Gaussian-based MinT methods to proportions, it does so at the cost of introducing systematic bias upon transformation back to the original scale.

\subsection{Generating and tilting convolutions under structural constraints.} \label{subsect:GTCSC}
To estimate the predictive density of the proportion $u_{n_{u},t}$, we adopt a sample-based convolution approach. In the case of proportions, the hierarchical structure imposes bounds on the support of the lower-level series. We therefore represent each bottom-level predictive distribution using a four-parameter Beta distribution whose support is defined by a known maximum contribution implied by the hierarchy, $w_{c}$, where $\sum_{c=1}^{m}w_{c} = 1$. This formulation ensures that the simulated bottom-level values satisfy the structural bounds while retaining flexibility in the shape of the predictive distributions. We model the estimated individual bottom-level forecast $\hat{b}_{c, t}$ using a 4-parameter Beta distribution with shape parameters $(\alpha_{c, t}, \beta_{c, t})$ and support $[0, w_{c}]$, such that
\begin{equation*}
    b_{c, t} \mid \alpha_{c, t}, \beta_{c, t}  \sim \text{Beta}(\alpha_{c, t}, \beta_{c, t}, 0, w_{c}).
\end{equation*}
Conditional on $b_{c, t}$, the density of the aggregated proportion, $u_{n_{u}, t}$, is then approximated via convolutions. Using Equation \eqref{eqn_contconv}, the density of $u_{n_{u},t}$ is expressed as,
\begin{align}
f_{n, t}(u_{n_{u}, t})
  &= \int_{0}^{w_{1}} \dots \int_{0}^{w_{m}}
    \bigg( \prod_{c=1}^{m-1} f_{c, t}(b_{c, t})   \bigg ) 
     f_{m, t}( u_{n_{u}, t}
       - \sum_{i=1}^{m-1} b_{i, t})
     \, db.
\end{align} 
A closed form solution for the convolution of Beta distributions does not exist, as such it is necessary to estimate the integral via Monte-Carlo integration. To do this, $R$ samples are generated from $m-1$ bottom series predictive four-parameter Beta distributions such that, 
\begin{align*}
    Q_{t+h}^{r} = \sum_{c=1}^{m-1} b_{c, t}^{r}, \quad r \in 1,\dots, R. 
\end{align*}
For a given value $z \in [0,1]$ of the target aggregation, the convolution is estimated by evaluating the predictive density of the remaining bottom-level series $m$ at the residual value $z - Q_{t}^{r}$ and averaging over the $R$ samples,
\begin{align*}
    \hat{f}_{n, t}(u_{n_{u}, t}) = \frac{1}{R}\sum_{r=1}^{R}f_{m, t}(z - Q_{t+h}^{r}) \approx f_{n, t}(u_{n_{u}, t}).
\end{align*}
Next, we apply an exponential tilting transformation to the convoluted density aligning the mean with the predictive mean of the top-level series. 

\paragraph{Exponential tilting of convoluted Beta densities.}  We assume that forecasts for the top-level series $u_{n_{u},t}$ and predictive mean from this series $\hat{u}_{n_{u}, t}$ are available. The convoluted density ($f_{n, t}(u_{n_{u}, t})$) is available numerically on a fine grid of values between 0 and 1. We therefore apply the exponential tilting transformation to the numerically estimated density to align its mean with the predictive mean of the top-level series. The tilted density $f^{*}_{n, t}$ is defined as,
\begin{equation}
    f^{*}_{n, t}(u_{n_{u}, t}) 
    = \frac{\text{exp}(\gamma_{n_{u}, t}u_{n_{u}, t})f_{n, t}(u_{n_{u}, t})}{\int_{0}^{1} \text{exp}(\gamma_{n_{u}, t}u_{n_{u}, t})f_{n, t}(u_{n_{u}, t}) \text{du}}
\end{equation}
The target of the tilting algorithm is to seek the optimal $\gamma_{n_{u}, t}$ that matches the mean of the convoluted density to the predictive mean of the top-level series. Because the convoluted density is available only numerically via $\hat{f}_{n, t}(u_{n_{u}, t})$, the tilting algorithm is approximated over the density grid, and a one-dimensional numerical root-finding algorithm is used to obtain $\gamma_{n_u, t}$. In our implementation, the root is obtained using the \texttt{uniroot} function in \textsf{R}. Further details of the estimation of the tilting parameter are provided in Appendix~\ref{appendix_tilting_details}.
Finally, the tilted coherent density can be sampled from, ensuring that the adjusted distribution preserves the original shape of $f^{*}_{n, t}(u_{n_{u}, t}) $ while shifting its mean to the target level $\hat{u}_{n_{u}, t}$. 
Where, the expectation of the top-level series tilted density is given by,
\begin{align}
    & E_{f_{n, t}^{*}}(u_{n_{u}, t}) = \int_{0}^{1} u_{n_{u}, t}f^{*}_{n, t}(u_{n_{u}, t})\text{du} \approx \hat{u}_{n_u, t}, \nonumber
\end{align}

\subsection{Simulation 2: Aggregating proportional forecasts} \label{sec:sim2}
\paragraph{Data description.}
The second simulation generates a hierarchical time-series of proportions within a two-level compositional structure, consisting of a single aggregate series and two bottom-level components 2500 observations. Due to the bounded and compositional nature of the data, the values across nodes are deterministically linked. As a result, only half of the full hierarchical tree must be explicitly simulated, since the remaining nodes can be obtained through deterministic aggregation or complementation. This reduced representation is sufficient for full reconciliation of the hierarchy. The bottom-level series $(b_{1, t}, b_{2, t})$ are constructed from auto-regressive latent processes on the logit scale with correlated Gaussian noise terms. Specifically, for $m = \{1,2\}$, $t = \{1,\dots,N\}$ and $N = 2500$, we define
\begin{equation}
    \begin{aligned}
    \text{logit}(b_{m,1}) &\sim \text{Normal}(0,1),\\
    \text{logit}(b_{m,t}) &= \kappa_{0,m} + \kappa_{1,m}\,b_{m,t-1} + \epsilon_{m, t}\\
    \end{aligned}
\end{equation}
We set the coefficients of the data generating process to be,
\begin{align}
& \boldsymbol{\kappa_{0}} = \begin{pmatrix}
0.85 \\
0.50
\end{pmatrix}, \quad \boldsymbol{\kappa_{1}}= \begin{pmatrix}
0.50\\
0.40
\end{pmatrix}, \quad
\boldsymbol{\epsilon}_{t}
\sim
\text{Normal}_2\!\left(
\begin{pmatrix}
0\\
0
\end{pmatrix},
\Sigma
\right), \nonumber \\
&\Sigma =
\begin{pmatrix}
(0.15)^2 & -0.6(0.15)^2 \\
-0.6(0.15)^2 & (0.15)^2
\end{pmatrix}. \nonumber
\end{align}
The bottom-level proportions are obtained by the inverse-logit transform. We define the mean proportion for the higher series node $u_{t}$ as the weighted sum of the lower–level series $b_{1, t}$ and $b_{2, t}$, $\bar{u}_{t} = 0.75b_{1, t}+ 0.25b_{2, t}$. To prevent a deterministic relationship between the nodes, we introduce noise via a precision parameter $\nu_u = 200$. Therefore, the larger the magnitude of $\nu_u$, the less noise induced by the sampling from the Beta distribution. Finally, we simulate draws from a Beta distribution,
\begin{equation}
    u_{t} \sim \text{Beta}\big(\nu_{u} \bar{u}_{t},\; \nu_{u}(1 - \bar{u}_{t})\big).
\end{equation}
The hierarchical composition of the final simulated dataset is given by, 
\begin{align*}
    \boldsymbol{y}_t = \begin{pmatrix} u_{t} \\ b_{1,t} \\ b_{2,t} \\ \end{pmatrix}
    = \begin{pmatrix} u_{t} \\ \boldsymbol{b}_t \end{pmatrix}.
\end{align*}
The data was split into a training and test set. The training set (2000 observations) across each node is used to build the models, while the test set (500 observations) is used to assess the predictive power and reliability of the hierarchical coherent forecast method. Thus, the coherent forecasts are evaluated on on their predictive performance of the test set.

\paragraph{Modelling the base forecasts.}  We assume that the modelling process for both $u_{t}$ and $b_{m,t}$ is the same. As such, we will just describe $b_{m,t}$. For node $b_{m,t}$ at time $t$, we assume the logit-transformed observation follows an ARMA($p,q$) process given by,
\begin{align*}
    \text{logit}(b_{m,t}) &= \eta_{m} + \sum_{p=1}^{P}\phi_{m,p}\,b_{m,t-p} + \epsilon_{m, t} +  \sum_{q=1}^{Q} \varphi_{m,q}\epsilon_{m, t-q}, \quad
    \epsilon_{m, t} \sim N(0, \sigma_{m}^2)
\end{align*}
where, $\eta_{m}$ is the node-specific intercept, $\boldsymbol{\phi}_{m}$ is the vector of $P$ autoregressive coefficients, and $\boldsymbol{\varphi}_{i}$ is the vector of $Q$ moving average coefficients. We estimate the parameters of the ARMA process using the \texttt{auto.arima} function of the \texttt{forecast} package in \textsf{R} \citep{Hyndman2025forecast}. The ARMA model is fitted on the logit scale. To construct the predictive distribution on the original proportion scale, the forecast mean and variance are transformed using the inverse-logit function and the delta method, respectively \citep{robinson2024deltamethod}. To generate samples from the four-parameter Beta predictive distribution, the forecast mean $\hat{b}_{m,t}$ and variance $\sigma^2_{b_{m,t}}$, expressed on the proportion scale, are used to obtain the corresponding Beta shape parameters $\alpha_{b_{m,t}}$ and $\beta_{b_{m,t}}$. The parameters are given by,
\begin{align}
    \alpha_{b_{m,t}}&= \left(\frac{1-\mu_{b_{m,t}}}{\sigma_{b_{m,t}}^2} - \frac{1}{\mu_{b_{m,t}}}\right)\mu_{b_{m,t}}^2, \\
    \beta_{b_{m,t}} &= \alpha_{b_{m,t}}\left(\frac{1}{\mu_{b_{m,t}}}-1\right).
\end{align}
where, $\mu_{b_{m,t}}=\hat{b}_{m,t}$ and $\sigma_{b_{m,t}} = \hat{\sigma}_{b_{m,t}}$.
The resulting four-parameter Beta distribution is then defined on the support determined by the structural bound of the bottom-level series,
\begin{equation}
    b_{m,t} \sim \operatorname{Beta} \left( \alpha_{b_{m,t}}, \beta_{b_{m,t}}, 0,w_m \right).
\end{equation}
We then follow the procedures described in section \ref{subsect:GTCSC} and Appendix \ref{appendix_tilting_details} to construct a joint density for the aggregate via convolutions and independently tilt this density towards the predictive mean of the top-series.

\paragraph{Using MinT with a logit link.}Since beta-distributed data are constrained to the unit interval, we first transform proportions using the logit link and then apply the MinT reconciliation method, employing a summation matrix that reflects the aggregation weights (0.75, 0.25) of the child series, and shrinkage estimator to ensure numerical stability of the covariance. The delta-method is used to transform the variance-covariance matrix of the residuals onto the logistic scale \citep{van2000asymptotic}. Reconciliation is performed on the transformed scale, and the resulting forecasts are subsequently back-transformed to the unit interval $[0,1]$. Unless otherwise stated, all case studies involving proportion data employ the logit shrinkage MinT reconciliation procedure.
 
\begin{figure}[H]
\centering
\includegraphics[width=15cm]{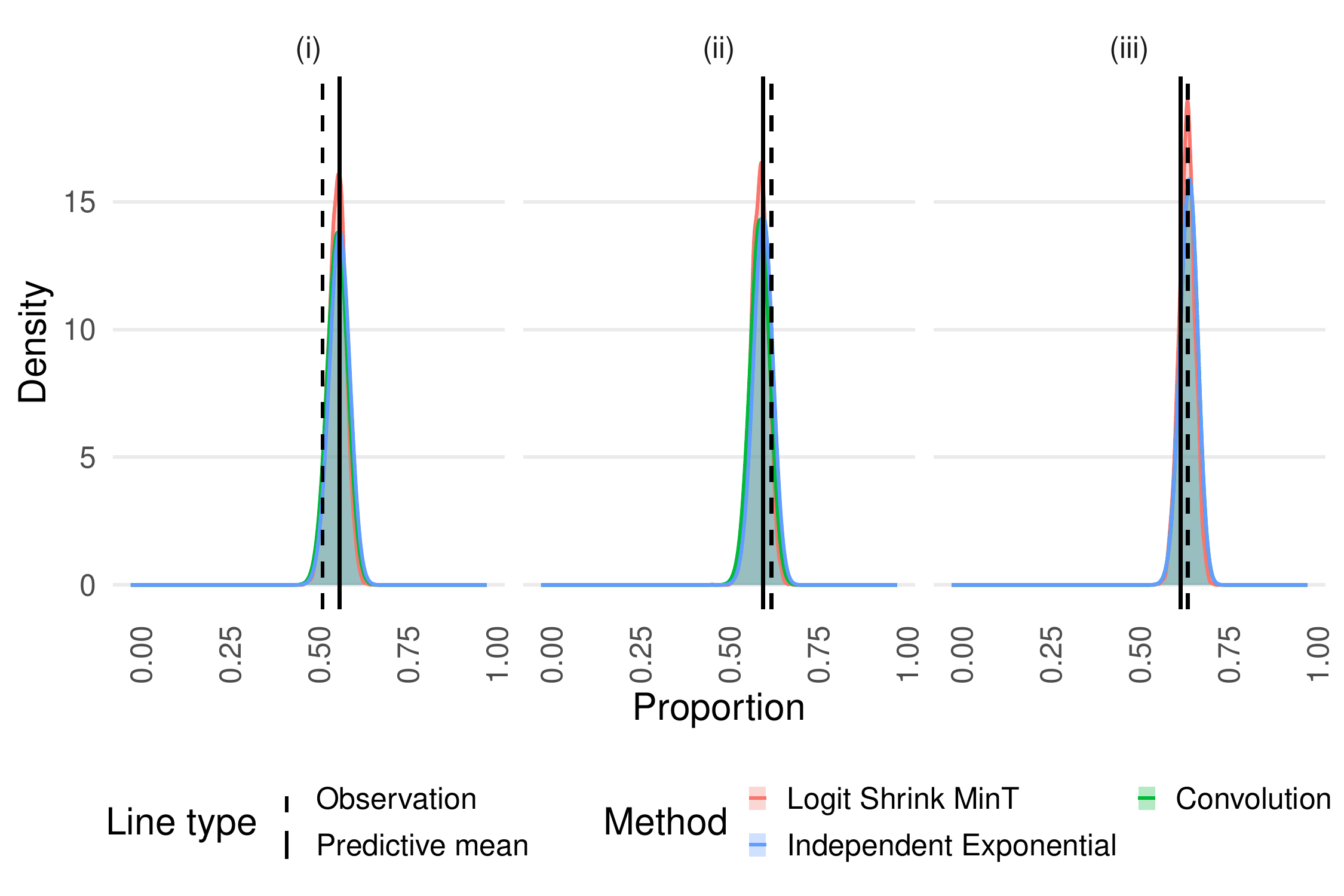}
\caption{Comparisons of the estimated predictive densities from each method are shown for selected validation observations from the simulated Beta datasets. In each panel, three test observations are displayed, chosen to span the range of the simulated data. Each method is represented by a distinct colour. The predictive mean for each method is indicated by a solid vertical line, while the observed value is shown by a dashed line. The Convolution method (green) represents densities estimated using convolution only; Convolution \& tilting (blue) represents densities estimated using convolution combined with Independent Exponential tilting; and Logit shrinkage MinT (red) represents the logit-scale shrinkage MinT reconciliation approach.}
\label{fig_density_beta_sim}
\end{figure}

\paragraph{Evaluation of the simulation study.}
To assess the performance of the proposed approach, we consider a simulated hierarchical Beta dataset with a two-level structure. The final 500 observations were withheld for validation, and one-step-ahead forecasts were generated sequentially using the corresponding validation observations. We first compare the marginal top-series forecast densities produced by three methods, a convolution, convolution with independent exponential tilting and logit-shrink MinT reconciliation (Figure \ref{fig_density_beta_sim}). \\
\\
In this example, all three methods perform similarly. The shape of the marginal distribution produced by both the convolution and independent tilted convolution overlap heavily. The peak of the logit-MinT density aligns with predictive mean, and has the highest peak overall. This is because the density of both convolution-based methods are evaluated using the entire predictive space  and therefore the value of the density is non-0, albeit extremely tiny, causing longer tails. Across the three methods, the logit–shrink MinT density exhibits the highest peak and the narrowest dispersion, indicating a more concentrated predictive distribution. This suggests that the logit–shrink MinT approach yields lower forecast uncertainty relative to the other two methods. \\
\\
Finally, we evaluate predictive accuracy using the Energy Score for the joint distribution of the hierarchy (AA, AB, and A) and the CRPS for the marginal distribution of the top series (Table \ref{tab:relative_to_LBMT_beta_demo}) relative to the performance of the logit-shrink MinT. When applying the Diebold–Mariano (DM) test, we conducted comparisons between the Convolution, Convolution with Independent Exponential tilting, and logit-scale MinT methods. At the 5\% significance level, we fail to reject the null hypothesis of equal predictive accuracy for all pairwise comparisons, indicating no statistically significant differences in performance between the methods across both datasets. In both metrics, the methods perform equally well. This aligns with the highly overlapping densities seen in Figure \ref{fig_density_beta_sim}.

% Place figure captions after the first paragraph in which they are cited.

\begin{table}[htbp]
\centering
\caption{Probabilistic forecast performance comparison relative to the lower-bounded MinT (LB MinT) method for different methods across two scoring rules (CRPS and Energy Score) using a test set of 500 observations. Lower values indicate better performance for the scoring rules. Pairwise Diebold-Mariano (DM) tests compare forecast accuracy between each reconciliation method and the LB MinT method, using the difference in Energy Scores of the joint distributions as the loss differential. The null hypothesis is equal predictive accuracy. Asterisks denote statistically significant differences from LB MinT according to the DM test (* $p<0.05$, ** $p<0.01$). In the method column, `Convolution' represents the density estimated using convolution only, `LB MinT' represents lower-bounded MinT reconciliation method, and `Convolutions with tilting' represents the density estimated using convolution and independent exponential tilting. Top marginal refers to the marginal distribution of the top series, while joint distribution refers to the joint distribution of the complete hierarchical series.}
\label{tab:relative_to_LBMT_beta_demo}
\small
\begin{tabular}{lrrrr}
\toprule
\multirow{2}{*}{Method} &
\multicolumn{1}{c}{\shortstack{Top \\ marginal}} &
\multicolumn{1}{c}{\shortstack{Joint \\ distribution}} &
\multicolumn{1}{c}{DM test} \\
\cmidrule(lr){2-2}
\cmidrule(lr){3-3}
\cmidrule(lr){4-4}
& CRPS & Energy & $p$-value \\
\midrule
LB MinT
& 1.000 & 1.000 & -- \\
Convolution
& 0.978  & 0.995 & 0.0612 \\
\shortstack{Convolutions \\ with tilting}
& 0.978 & 0.997 & 0.418 \\
\bottomrule
\end{tabular}
\end{table}

\subsection{Application to Australian infant death prediction}
\begin{figure}[H]
\centering
\includegraphics[width=14cm]{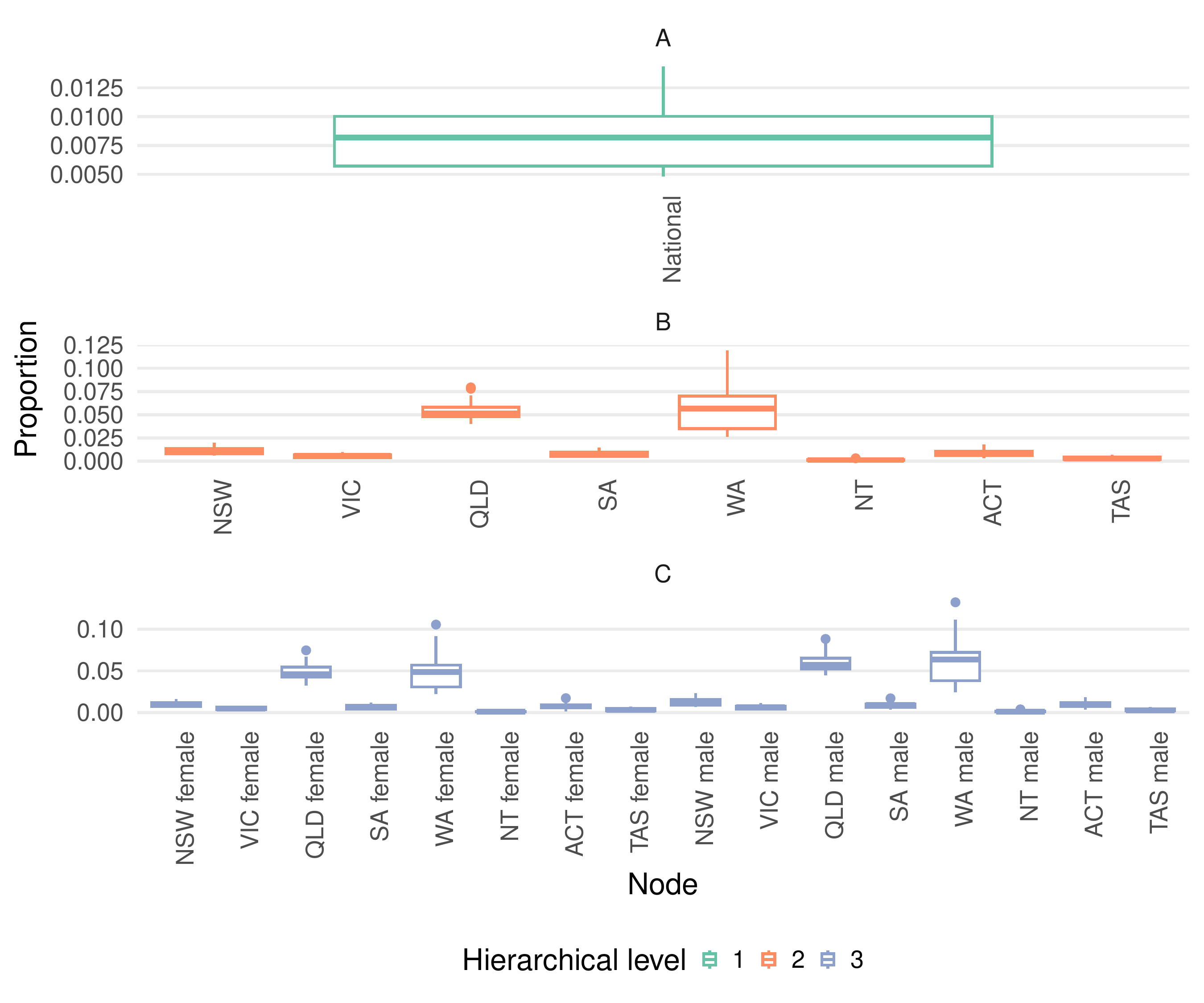}
\caption{Boxplots of observed infant mortality proportions across the hierarchical structure, including the national level, state-level aggregates, and sex-specific disaggregations. The distribution at each node highlights variability and differences in scale across levels. }
\label{fig_boxplot_infantdeaths}
\end{figure}

\paragraph{Data description.} 
The \texttt{infantdeaths} dataset from the \texttt{hts} package in R contains annual counts of infant deaths in Australia from 1933 to 2003, disaggregated by sex (male and female) and by state/territory \citep{hts}. The data are structured as a hierarchical time series with three levels: the bottom level represents deaths by state and sex, the intermediate level aggregates deaths by state, and the top level corresponds to total deaths across Australia. For the purposes of this illustration, we consider the proportion of births resulting in death over time at the national-, state-, and state-sex levels. We accessed the annual number of births via the Australian Bureau of Statistics website \citep{ABS}.  As the number of births disaggregated by sex was not available from the Australian Bureau of Statistics at the time of writing, we assumed that the number of state-level births disaggregated by the average sex-ratio (female:male) of 100:105 births as given by Australian Bureau of Statistics website \citep{ABS}. The resulting proportions range from 0.04\% to 13\% approximately, with higher variability at lower disaggregation levels (Figure \ref{fig_boxplot_infantdeaths}). The hierarchical relationships are defined as
\begin{align*}
    u_{n_{u}, t} &= \sum_{k=1}^{n_{u}-1} w_{1, k} u_{k,t}, \\
    u_{k, t} &= \sum_{c \in \mathcal{D}_{k}} w_{2,c}\, b_{c,t}, \quad k = 1, \ldots, n_{u}-1,
\end{align*}
where $u_{n_{u},t}$ denotes the top-level (national) series at time $t$, and $n_{u}$ is the total number of aggregated nodes in the hierarchy ($n_{u} = 9$). The bottom-level series are denoted by $b_{c,t}$, with $c = (1, \ldots, 16)$. The set $\mathcal{D}_{k}$ indexes the child nodes associated with aggregated node $k$. The weights $w_{1,c}$ and $w_{2,k}$ represent aggregation weights for constructing the top and intermediate levels, respectively. In this application, each intermediate node $u_{k,t}$ corresponds to a state-level series and is constructed from $|\mathcal{D}_{k}| = 2$ bottom-level series (male and female components).
% Place figure captions after the first paragraph in which they are cited.

Sex- and state-specific weights are used later as support constraints in the 4-parameter Beta distribution, enabling child distributions to aggregate to the parent distributions, while respecting the support constraints of the parent node. For the aggregation of the bottom series sex-specific nodes to mid-tier state-level nodes, we assume a fixed sex birth ratio of 100:105 (female:male). Then, the weights are defined as
\begin{align*}
w_{1,c} = \frac{P_c}{\sum_{c=1}^2 P_c}
\end{align*}
where $P_c$ denotes the average number of births by sex $c$. For the aggregation of state-level nodes to the top-series national node, weights are based on each state’s average share of national births, which is given by,
\begin{align*}
w_{2,k} &= \frac{\sum_{t=1}^{T} O_{k,t}}{\sum_{t=1}^{T} \sum_{k=1}^{K} O_{k,t}}.
\end{align*}
where $O_{k,t}$ denotes the number of births in state $k$ at time $t$.
\paragraph{Modelling the base forecasts.} We follow the same approach to modelling the proportions using an ARMA model and logit link as described in Simulation 2, section \ref{sec:sim2}. We then estimated and tilted the joint densities of the coherent series using the approach described in section \ref{subsect:GTCSC} and Appendix \ref{appendix_tilting_details}. The data was split into a training and test set. The test set of this data consisted of the final 11 years of observations across the three tiers of the hierarchical tree (11 years x 26 nodes). The training set (observations before 1993) is used to build the models, while the test set (observations on or after 1993) is used to assess the predictive power and reliability of the hierarchical coherent forecast method. Thus, the coherent forecasts are evaluated on on their predictive performance of the test set.

\paragraph{Using MinT via a logit link.} We follow the same approach to reconciling proportions using MinT via a logit link as described in Simulation 2, section \ref{sec:sim2}.

\begin{figure}[H]
\centering
\includegraphics[width=17cm]{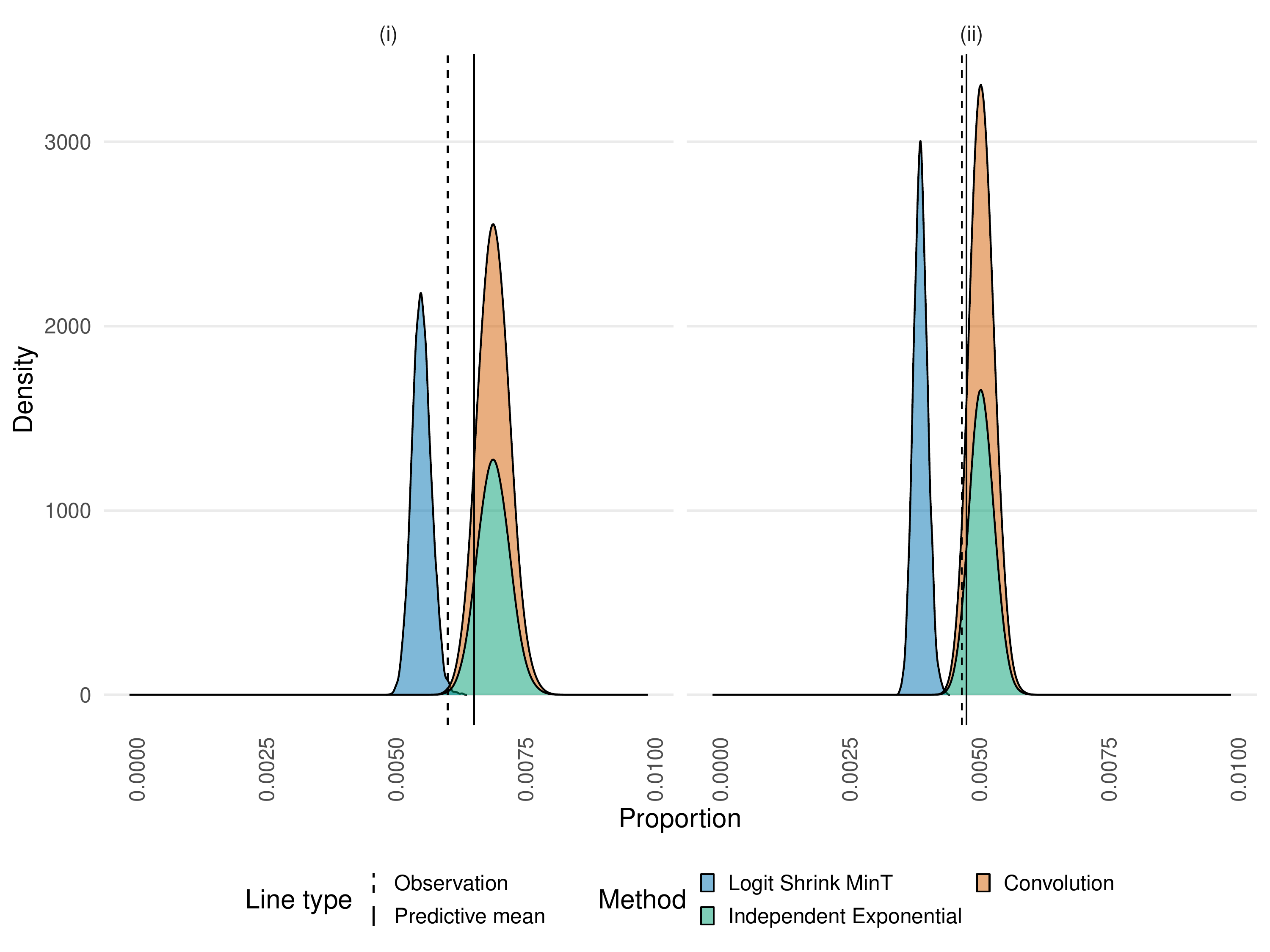}
\caption{The estimated densities for two years of validation data at the national level across the methods considered. Each colour represents a different density estimation method. 'Convolution' in green represents the density estimated using a convolution only, 'Independent Exponential' in blue represents the density estimated using a convolution and exponential tilting with the tilting parameters estimated independently. `Logit shrinkage MinT` in red represents the logit-scale shrinkage MinT reconciliation approach. The predictive mean for the national-level test set observation is given by the vertical line. For clarity, the range of this figure is truncated, however the density is evaluated across [0, 0.25] for convolution-based methods.}
\label{fig_density_infants_national}
\end{figure}

\paragraph{Evaluation of the Australian infant deaths dataset.} To evaluate the proposed approach, we compare the coherent proportions with a logit-scale shrinkage MinT reconciliation approach. The proportions in the validation set are extremely small and lie close to the boundary of the unit interval, making them inherently more challenging to estimate and reconcile than proportions located in the interior of the support. \\
\\
This behaviour is illustrated in Figure \ref{fig_density_infants_national}, which presents the predictive marginal densities at the national level for two selected years. For clarity, the displayed range is truncated, although the density for the convolution-based methods is evaluated over $[0,0.25]$. The predictive means in these examples are close to zero. Near the boundary of the unit interval, the density of a Beta distribution can become highly skewed, while the logistic transformation is in its asymptotic region, such that changes on the logit scale correspond to only small changes on the probability scale. The figure also illustrates that, when the difference between the predictive mean and the mean of the convoluted density is negligible, the first-moment adjustment from exponential tilting is correspondingly small. Consequently, the independently tilted densities closely overlap the original convoluted densities, with the primary difference being somewhat thicker tails. In contrast, the logit-scale shrinkage MinT approach produces a highly concentrated predictive density but systematically underestimates the predictive mean. This suggests that the shrinkage reconciliation is over-adjusting the forecasts towards coherence in this example, resulting in both the predictive mean and the observed value being poorly represented. \\
\\
The scoring results in Table \ref{tab:relative_to_LBMT_beta_infants} provide further evidence of this difference. Unlike LB MinT, the proposed convolution-based methods do not modify the bottom-level series. This has a substantial effect on the Energy Score of the resulting joint distributions: for both convolution-based methods, the joint Energy Score is approximately seven times larger than that of LB MinT. This difference appears to be driven primarily by the bottom-level series, as the Energy Scores calculated using only the middle- and national-level nodes are lower than those obtained under LB MinT. At the national level, however, the marginal CRPS is substantially lower for both convolution-based methods (0.282 and 0.284, respectively), providing further evidence that the LB MinT forecasts are being over-adjusted in this example. At the state level, the marginal CRPS values are broadly comparable with those of LB MinT, although performance varies across states. For example, the convolution-based methods provide a poorer approximation for VIC but an improvement for QLD. Finally, the difference in joint Energy Scores is statistically significant, leading us to reject the null hypothesis of equal forecast accuracy between each of the convolution-based methods and LB MinT. Thus, although the convolution-based approaches provide competitive or improved marginal forecasts at the middle and national levels, their failure to adjust the bottom-level distributions results in substantially poorer performance under the joint Energy Score in this example.

% Place figure captions after the first paragraph in which they are cited.

\begin{table}[htbp]
\centering
\caption{Probabilistic forecast performance comparison relative to the lower-bounded MinT (LB MinT) method for different methods across two scoring rules (CRPS and Energy Score) using the test set of 26 years of data across the complete series. Lower values indicate better relative performance for the scoring rules. Pairwise Diebold-Mariano (DM) tests compare forecast accuracy between each reconciliation method and the LB MinT method, using the difference in Energy Scores of the joint distributions as the loss differential. The null hypothesis is equal predictive accuracy. Asterisks denote statistically significant differences from LB MinT according to the DM test (* $p<0.05$, ** $p<0.01$). In the method column, `Convolution' represents the density estimated using convolution only, `LB MinT' represents lower-bounded MinT reconciliation method, and `Convolutions with tilting' represents the density estimated using convolution and independent exponential tilting. `Top marginal' refers to the marginal distribution of the top series, `Mid marginal' refers to the marginal distribution of the middle series, `Joint distribution' refers to the joint distribution of the complete hierarchical series and `Reduced joint distribution' refers to the joint distribution of the top- and mid- hierarchical series only.}
\label{tab:relative_to_LBMT_beta_infants}
\small
\resizebox{\textwidth}{!}{
\begin{tabular}{lrrrrrrrrrrrr}
\toprule
\multirow{2}{*}{Method} &
\multicolumn{1}{c}{\shortstack{Top \\ marginal}} &
\multicolumn{1}{c}{\shortstack{Joint \\ distribution}} &
\multicolumn{1}{c}{\shortstack{Reduced joint \\ distribution}} &
\multicolumn{8}{c}{\shortstack{Mid marginals}} &
\multicolumn{1}{c}{DM test} \\
\cmidrule(lr){2-2}
\cmidrule(lr){3-3}
\cmidrule(lr){4-4}
\cmidrule(lr){5-12}
\cmidrule(lr){13-13}
& CRPS & Energy & Energy
& NSW & VIC & QLD & SA & WA & NT & ACT & TAS
& $p$-value \\
\midrule
LB MinT
& 1.000 & 1.000 & 1.000
& 1.000 & 1.000 & 1.000 & 1.000
& 1.000 & 1.000 & 1.000 & 1.000
& -- \\

Convolution
& 0.282 & 7.060 & 0.905
& 0.937 & 1.210 & 0.890 & 0.980
& 1.010 & 0.733 & 0.902 & 0.992
& $2.2^{-10}**$ \\

\shortstack{Independent \\ exponential}
& 0.284 & 7.060 & 0.914
& 0.927 & 1.200 & 0.898 & 0.964
& 1.000 & 0.737 & 0.922 & 1.000
& $2.2^{-10}**$ \\
\bottomrule
\end{tabular}
}
\end{table}

\section{Discussion}  \label{discussion}
Distributional forecasting is an area of growing interest. As Kolassa (2023) argues, the focus in hierarchical forecasting needs to shift from point forecasts to coherent density forecasting. That is, estimating predictive joint densities in a coherent manner and using them to derive optimal point forecasts, rather than merely ensuring coherence of individual point forecasts. Our proposed method, which combines convolutions with exponential tilting, produces coherent distributional forecasts for bounded or non-negative discrete and continuous hierarchical series. In this paper, we focus on linear hierarchical coherent forecasting, where hierarchical relationships are defined by linear sums of nodes. For future research, we aim to extend this framework to non-linear and sequential hierarchical relationships. Additionally, while this study considers relatively small hierarchies (fewer than 100 nodes in the complete series), evaluating the efficiency and performance of our method on larger hierarchical series is a natural next step. A limitation of the proposed method is that the tilting step relies on confidence in the predictive means of the base regression models. This approach assumes correctly specified models without mis-specification. While we employed relatively simple models to generate base forecasts, adopting more complex models could improve predictive accuracy. \\
\\
Nevertheless, our goal here was to demonstrate the strengths of the tilting-based reconciliation method. We therefore recommend that users carefully validate base forecasts before applying the post-hoc coherency adjustment. The key contribution of this work is that, to the best of our knowledge, it is the first method specifically designed to reconcile proportions without transforming the bounded nature of the data. Moreover, it handles both continuous and discrete distributions, advancing the area of coherent hierarchical forecasting. Our method performs favourably compared to other established approaches for discrete and bounded forecast reconciliation, highlighting its potential as a robust tool for practical applications.\\
\\
To conclude, our approach to distributional forecasting leverages convolutions combined with exponential tilting to generate coherent hierarchical forecasts for both discrete and continuous bounded and constrained distributions. We have demonstrated the utility of this method across multiple examples involving Beta and Poisson-distributed data, showing that it consistently matches existing state-of-the-art reconciliation methods. These results highlight the flexibility and effectiveness of our approach for producing coherent predictive distributions in hierarchical settings.

\small
\bibliography{references.bib}

\pagebreak

\appendix
\section{Estimation of $\gamma_{i,t}$ for density tilting} 
\label{appendix_tilting_details}
We estimate the tilted density using the moment generating function (MGF) of the original density $f_{i,t}$ together with a saddlepoint approximation \citep{Butler2010_ExponentialFamilies}. Let the MGF of $f_{i,t}$ be denoted by $M(\gamma_{i,t})$, and let $\gamma_{i,t}$ be the tilting parameter to be optimised. The MGF and its log-transform, the cumulant generating function, are defined as,  
\begin{align}
    M(\gamma_{i,t}) &= \int \exp(\gamma_{i,t} y_{i,t}) f_{i,t}(y_{i,t}) ,dy_{i,t}, \\
    K(\gamma_{i,t}) &= \log(M(\gamma_{i,t})).
\end{align}
The tilted density $f_{i,t}^{*}(y_{i,t})$ then inherits its moments from the derivatives of the cumulant generating function $K(\gamma_{i,t})$. In our approach, we utilise the first moment and link it $K(\gamma_{i,t})$, such that,
\begin{align}
    \frac{dK(\gamma_{i,t})}{d\gamma_{i,t}} &= E_{f_{i,t}^{*}}(y_{i,t}) \\
                                           &= \frac{d}{d\gamma_{i,t}} \log\left( \int \exp(\gamma_{i,t} y_{i,t}) f_{i,t}(y_{i,t}) ,dy_{i,t}\right) \\
                                           &= \frac{\int y_{i,t}\exp(\gamma_{i,t} y_{i,t}) f_{i,t}(y_{i,t}) ,dy_{i,t}}
                                           {\int \exp(\gamma_{i,t} y_{i,t}) f_{i,t}(y_{i,t}) ,dy_{i,t}}.
\end{align}
The derivative $\frac{dK(\gamma_{i,t})}{d\gamma_{i,t}}$ may be numerically approximated such that, 
\begin{align}
    E_{f_{i,t}^{*}}(y_{i,t}) &\approx \frac{\sum_{j=1}^{N} y_{i,t}^{(j)}\delta_j \exp(\gamma_{i,t} y_{i,t}^{(j)}) f_{i,t}(y_{i,t}^{(j)})}
                                       {\sum_{j=1}^{N} \delta_j \exp(\gamma_{i,t} y_{i,t}^{(j)}) f_{i,t}(y_{i,t}^{(j)})},
\end{align}
where $\delta_j$ is the grid spacing between adjacent points in $y_{i,t}$ used to compute the convolution.
If this approximation is done via Monte Carlo, the same samples from the density $f_{i,t}(y_{i,t})$ can be used when evaluating different values of $\gamma_{i,t}$, making the algorithm computationally efficient. Finally, the density of the exponentially tilted distribution is given by,
\begin{align}
    f_{i,t}^{*}(y_{i,t}) = \frac{\exp(\gamma_{i,t} y_{i,t}) f_{i,t}(y_{i,t})}{\int \exp(\gamma_{i,t} y_{i,t})f_{i,t}(y_{i,t}) ,dy_{i,t}}.
\label{eqnExptilt}
\end{align}
Multiplying by $\text{exp}(\gamma_{i,t} g(y_{i,t}))$ increases the density of points in the sample space for which $\gamma_{i,t} g(y_{i,t})$ is the largest relative to those where $\gamma_{i,t} g(y_{i,t})$ is smaller. The moment-condition is solved numerically for $\gamma_{i,t}$ (e.g., via a root-finding algorithm such as \texttt{uniroot} in \textsf{R} \citep{R-base}). If the constraint does not admit a solution, for example when the required moment lies outside the range of the tilted distribution, then the distribution cannot be tilted without violating the probability structure. In such cases, we set $\gamma_{i,t}$=0. Therefore, no tilting is applied and the original (non-tilted) density is retained.

\section{Algorithms}
The algorithm takes as input the vector of Poisson regression forecasts across the complete series for a given time $t$, 
$\hat{\mathbf{y}}_{t} = (\hat{\lambda}_{1,t}, \dots, \hat{\lambda}_{N,t})$.
It applies exponential tilting to enforce theoretical mean restrictions, and returns the tilted predictive density of the top series, $\check{f}_{N}(u_{t})$.

\begin{algorithm}[H]
\caption{Coherent forecasting via convolution and exponential tilting (Poisson)}
\begin{algorithmic}
\For{$t = 1, \dots, T$}

    \Statex \textbf{Convolution step:}
    
    \State The aggregate PMF is given by
    \State $f_{N}(u_t = l) = \frac{exp(-{\sum_{m=1}^M \hat{\lambda}_{m,t})}(\sum_{m=1}^M \hat{\lambda}_{m,t})^{l}}{l!}$
    
    \State For Poisson marginals:
    \State $u_t \sim \text{Poisson}\left(\sum_{m=1}^M \hat{\lambda}_{m,t}\right)$

    \Statex \textbf{Exponential tilting:}
    
    \State $M(\gamma_t) = \sum_{l=0}^L e^{\gamma_t l} f_{N}(u_t)$
    
    \State Solve for $\gamma_t$ such that,
    \State $\displaystyle \frac{\sum_{l=0}^L l e^{\gamma_t l} f_{N}(u_t)}{\sum_{l=0}^L e^{\gamma_t l} f_{N}(u_t)} = \hat{\lambda}_{N,t}$\\
    
    \State The tilted mass function is given by:
    \State $\check{f}_{N}(u_t = l \mid \gamma_t) = \frac{ e^{\gamma_t l}\, \dfrac{\sum_{m=1}^M \hat{\lambda}_{m,t}^{\,l}}{l!}}{\sum_{l=0}^{L} e^{\gamma_t l}\, \dfrac{\sum_{m=1}^M \hat{\lambda}_{m,t}^{\,l}}{l!}}$.

\EndFor
\end{algorithmic}
\end{algorithm}

\subsection{Coherent forecasting via convolution and exponential tilting (Beta)}
The algorithm takes as input base forecast proportions $\hat{\boldsymbol{y}}_{t} = (\hat{p}_{1,t}, \dots, \hat{p}_{N,t})$ and associated shape and rate parameters, $\hat{\boldsymbol{\alpha}}_{t}$, $\hat{\boldsymbol{\beta}}_{t}$, and returns an estimate of the reconciled predictive density of the aggregate $u_{N,t} = \sum_{i=1}^m w_{i}p_{i,t}$ over the unit interval $(0,1)$, where $\{w_{i}\}$ are known weights and $p_{m,t} \in [0,1]$. 

\begin{algorithm}[H]
\caption{Coherent forecasting via convolution and exponential tilting (Beta)}
\begin{algorithmic}
\For{$t = 1, \dots, T$}

    \Statex \textbf{Convolution step:}
    
    \State The aggregate density is
    \State
    \begin{align*} 
        f_{N}(z)  = \int_{\prod_{m=1}^{M-1}[0,w_{m}]} & \prod_{m=1}^{M-1} f_m(b_{m,t} \mid \hat{\alpha}_{m,t}, \hat{\beta}_{m,t}, 0, w_{m}) \\ 
        & \times f_M\Big(z - \sum_{m=1}^{M-1} b_{m,t} \mid \hat{\alpha}_{M,t}, \hat{\beta}_{M,t}, 0, w_M \Big) db_1 \cdots db_M; 
    \end{align*} 

    \Statex \textbf{Exponential tilting:}
    
    \State $M_t(\gamma_t) = \int_0^1 e^{\gamma_t z} f_{N}(z)\,dz$
    \State Solve for $\gamma_t$ such that
    \State $\displaystyle \frac{\int_0^1 z e^{\gamma_t z} f_{N}(z)\,dz}{\int_0^1 e^{\gamma_t z} f_{N}(z)\,dz} = \hat{p}_{N,t}$.\\

    \State The tilted mass function is given by:
    \State $\check{f}_{N}(z, \gamma_t) = \frac{e^{\gamma_t z} f_{N}(z)}{M_t(\gamma_t)}$

\EndFor
\end{algorithmic}
\end{algorithm}

\section{Inverse-logit transformation of MinT: Implications on the assumption of unbiased predictors}
\label{InvLogit_MinT_Bias_Proof}
A key assumption of the MinT algorithm is the assumption of unbiased estimators \citep{Wickramasuriya2019}. This assumption is made based on the understanding that the MinT algorithm utilises a linear transformation of variables from child to parent nodes. Here, we consider what happens to the assumption of unbiased predictors when a non-linear transformation of variables is considered. To begin, let, 
\begin{align*}
   \hat{\epsilon}_{T}(h) = y_{T+h} - \hat{y}_{T}(h)
\end{align*}
be the h-step ahead conditionally stationary base forecast errors where $E(\hat{\epsilon}_{T}(h) \mid \boldsymbol{\mathcal{I}}_{T})=0$ and $\boldsymbol{\mathcal{I}}_{T}$ is the observed data $ \{ y_{1}, \dots y_{T} \}$ available up to time $T$. It is possible to decompose the mean squared error into components of bias, variance and irreducible error, such that,
\begin{align*}
   E(\hat{\epsilon}_{T}(h)^2) = \Big( E[\hat{y}_{T}(h)] -  y_{T+h} \Big)^2 + E\Big [ (\hat{y}_{T}(h) - E[\hat{y}_{T}(h)]  )^2 \Big ] + \sigma^2
\end{align*}
where the first term captures the bias, the second term captures the variance and $\sigma$ captures the irreducible error. Here, we consider the consequences of reconciling on the logit-scale and how back-transformation to the [0,1] scale affects the bias assumptions of MinT.
Let $Z$ be an estimator on the logit scale, such that 
\begin{align*}
    Z \sim N(\mu_z, \sigma_z^2)
\end{align*}
Let P be the target of interest, such that $p = \frac{1}{1+exp^{-z}}$, and $\hat{p} = g(\hat{z})$, where g is the inverse of the logit function.
First, we begin with an approximation of $g(\hat{z})$ around $\mu_{z}$ using a Taylor Series Expansion \citep{casella2002}, 
\begin{align*}
 g(\hat{z}) = g(\mu_{z}) - g'(\mu_{z})( \hat{z} - \mu_{z}) + \frac{1}{2}g''(\mu_{z})( \hat{z} - \mu_{z})^2
\end{align*}
We transform the variance of $Z$ using an inverse-logit transformation and the Delta Method \citep{casella2002}, where
\begin{align*}
    Var(g(\hat{z})) & \approx [ g'(\mu_{z})]^2\sigma_z^2, \\
    & = [ g(\mu_{z})(1 - g(\mu_{z}))]^2\sigma_z^2.
\end{align*}
The estimated bias is defined as, 
\begin{align*}
    \text{Bias}(g(\hat{z})) & = E(g(\hat{z})) - g(z), \\
     &=  g(\mu_{z}) - g'(\mu_{z}) E( \hat{z} - \mu_{z}) + \frac{1}{2}g''(\mu_{z})E( \hat{z} - \mu_{z})^2 - g(z), \\
     &=  \frac{1}{2}g''(\mu_{z})E( \hat{z} - \mu_{z})^2, \\
     & = \frac{1}{2}g''(\mu_{z}) \sigma_z^2.
\end{align*}
The mean squared error of the inverse-logit reconciled forecasts can be decomposed into it's variance and bias contributions by, $MSE(\hat{p}) = Var(\hat{p}) + Bias(\hat{p})^2$ \citep{casella2002}. Therefore, 
\begin{align*}
    \hat{\epsilon}_{p} &= \hat{p} - p, \\
    E(\hat{\epsilon}_{p}^2) &= \Big[g(\mu_{z})(1 - g(\mu_{z}))\Big]^2\sigma_z^2 + \frac{1}{4} \Big[ g''(\mu_{z})\Big]^2   \sigma_z^4 
\end{align*}
As the variance of the logit-transformed series $Z$, $\sigma_z$, is non-zero, we can conclude that the inverse-logit transformation will induce a bias in the reconciled estimates as $\sigma_z >0$.

\bibliographystyle{plainnat}

\end{document}